\documentclass[aps,prb,reprint,superscriptaddress,secnumarabic]{revtex4-2}

\usepackage{graphicx}
\usepackage[utf8]{inputenc}
\usepackage{float}
\usepackage{comment}
\usepackage{amsmath}

\usepackage{xcolor}

\begin{document}


\title{Temperature-dependent spin-resolved electronic structure of EuO thin films}

\author{Tristan Heider}
\affiliation{Peter Gr{\"u}nberg Institut (PGI-6), Forschungszentrum J{\"u}lich GmbH,
52428 J{\"u}lich, Germany}
\affiliation{Fakult{\"a}t f{\"u}r Physik, Universit{\"a}t Duisburg-Essen, 47048 Duisburg, Germany}

\author{Timm Gerber}
\affiliation{Peter Gr{\"u}nberg Institut (PGI-6), Forschungszentrum J{\"u}lich GmbH,
52428 J{\"u}lich, Germany}

\author{Okan K\"oksal}
\affiliation{Fakult{\"a}t f{\"u}r Physik, Universit{\"a}t Duisburg-Essen, 47048 Duisburg, Germany}

\author{Markus Eschbach}
\affiliation{Peter Gr{\"u}nberg Institut (PGI-6), Forschungszentrum J{\"u}lich GmbH,
52428 J{\"u}lich, Germany}
\author{Ewa M\l y\'{n}czak}
\affiliation{Peter Gr{\"u}nberg Institut (PGI-6), Forschungszentrum J{\"u}lich GmbH,
52428 J{\"u}lich, Germany}
\affiliation{J. Haber Institute of Catalysis and Surface Chemistry, Polish Academy of Sciences, 30-239 Krakow, Poland}

\author{Patrick L{\"o}mker}
\affiliation{Peter Gr{\"u}nberg Institut (PGI-6), Forschungszentrum J{\"u}lich GmbH,
52428 J{\"u}lich, Germany}
\affiliation{Photon Science, Deutsches Elektronen-Synchrotron DESY, 22607 Hamburg, Germany}
\author{Pika Gospodaric}
\author{Mathias Gehlmann}
\author{Moritz Pl\"otzing}
\affiliation{Peter Gr{\"u}nberg Institut (PGI-6), Forschungszentrum J{\"u}lich GmbH,
52428 J{\"u}lich, Germany}

\author{Rossitza Pentcheva}
\affiliation{Fakult{\"a}t f{\"u}r Physik, Universit{\"a}t Duisburg-Essen, 47048 Duisburg, Germany}

\author{Lukasz Plucinski}
\email{l.plucinski@fz-juelich.de}
\affiliation{Peter Gr{\"u}nberg Institut (PGI-6), Forschungszentrum J{\"u}lich GmbH,
52428 J{\"u}lich, Germany}
\affiliation{Fakult{\"a}t f{\"u}r Physik, Universit{\"a}t Duisburg-Essen, 47048 Duisburg, Germany}

\author{Claus M. Schneider}
\affiliation{Peter Gr{\"u}nberg Institut (PGI-6), Forschungszentrum J{\"u}lich GmbH,
52428 J{\"u}lich, Germany}
\affiliation{Fakult{\"a}t f{\"u}r Physik, Universit{\"a}t Duisburg-Essen, 47048 Duisburg, Germany}
\affiliation{Physics Department, University of California, Davis, CA 95616, USA}

\author{Martina M{\"u}ller}
\affiliation{Peter Gr{\"u}nberg Institut (PGI-6), Forschungszentrum J{\"u}lich GmbH,
52428 J{\"u}lich, Germany}
\affiliation{Fachbereich Physik, Universit{\"a}t Konstanz, 78464 Konstanz, Germany}


\date{\today}

\begin{abstract}
The electronic structure of the ferromagnetic semiconductor EuO is investigated by means of spin- and angle-resolved photoemission spectroscopy (spin-ARPES) and density functional theory. EuO exhibits unique properties of hosting both weakly-dispersive nearly fully polarized Eu $4f$ bands, as well as O $2p$ levels indirectly exchange-split by the interaction with Eu nearest neighbors. Our temperature-dependent spin-ARPES data directly demonstrates the exchange splitting in O $2p$ and its vanishing at the Curie temperature.  Our calculations with a Hubbard $U$ term reveal a complex nature of the local exchange splitting on the oxygen site and in conduction bands. We discuss the mechanisms of the indirect exchange in the O 2p levels by analyzing orbital-resolved band characters in ferromagnetic and antiferromagnetic phases. The directional effects due to spin-orbit coupling are predicted theoretically to be significant in particular in the Eu 4f band manifold. The analysis of the shape of spin-resolved spectra in the Eu $4f$ spectral region reveals signatures of hybridization with O $2p$, in agreement with the theoretical predictions. We also analyze spectral changes in the spin-integrated spectra throughout the Curie temperature and demonstrate they derive from both the magnetic phase transition and effects due to sample aging, unavoidable for this highly reactive material.
\end{abstract}


\maketitle
\section{Introduction}

The nature of the ferromagnetic phase transition has been one of the fundamental issues in condensed matter physics \cite{Stoehr2006,Kuebler2009}. Its theoretical treatment is often based on the framework of local magnetic moments (LMs), which is particularly suitable to model $4f$ compounds. LMs are all aligned parallel in the ferromagnetic (FM) phase well below the Curie temperature ($T_C$), fully disordered in the paramagnetic phase \cite{Stoehr2006, Heine1990, Heine1988, Kuebler2009}, and intermediate temperature range can be described by an increased occupation of various magnon modes. This framework has been used to model numerous important magnetic phenomena such as the nature of the magnetic order in dilute magnetic topological insulators \cite{Rosenberg2012,Peixoto2020}, the surface ferromagnetism in novel $4f$ compounds \cite{Chikina2014,Schulz2019}, and the short-range order in itinerant magnets \cite{Heine1990,Melnikov2019,Antropov2005,Kuebler2009}.


In this work, we are studying the electronic structure of EuO across the ferromagnetic-paramagnetic phase transition. We have chosen EuO as a prototypical Heisenberg FM, where Eu $4f$ orbitals can be treated as LMs. EuO is offering a benefit of an easily accessible Curie temperature ($T_C\sim 70 K$).
The LMs in EuO are Eu $4f$ which exhibit $\sim 10$ eV (on site) exchange splitting which is therefore virtually temperature independent. This indirectly leads to much smaller exchange splitting also in bands derived from other orbitals, and for Eu $5d$ has been described within the $d-f$ model \cite{Schiller2000,Schiller2001c}. Similar model has been recently used to establish the temperature-dependent exchange splitting in S $3p$ levels of EuS \cite{Fedorov2021}, a material similar to EuO but having a much smaller $T_C = 16.5$K.

We employ spin-resolved version of angle-resolved photoemission (spin-ARPES) to get access to the spin-polarized electronic structure of EuO up to $T_C$. Spin-ARPES has been the method of choice for studying the band structure of magnetic materials \cite{Osterwalder2006, Plucinski2013, Eich2016}, however, we are only aware of a single report on spin-resolved photoemission on EuO: in an angle-integrated study it was reported that the spectral polarization is temperature dependent and vanishes at $T_C$ \cite{Lee2007}. In the case of spin-integrated ARPES, a recent synchrotron study on Gd-doped EuO has reported for the first time results where clear dispersions in the O $2p$ manifold are present in experimental $E(k)$ maps \cite{Riley2018}. Earlier work by Miyazaki et al. \cite{Miyazaki2009} reported a vanishing of exchange splitting in O $2p$ at $T_C$, based on the analysis of the second derivative of ARPES spectra at $\Gamma$ and $X$ points of the Brillouin zone (BZ). Furthermore, the well-established red shift of the optical absorption edge \cite{Freiser1968,Busch1967,Prinz2016}, the temperature-dependent resistivity of EuO tunnel junctions \cite{Moodera2007, Mueller2009}, and the metal-insulator transition (MIT) in Eu-rich samples \cite{Steeneken2002b, Mairoser2010} have been ascribed to the vanishing of the exchange splitting in the Eu $5d$-derived conduction band minimum (CBM) at $T_C \simeq 70$ K \cite{Schiller2001c}. Previously mentioned work on EuS \cite{Fedorov2021} has addressed the spectral renormalization throughout $T_C$ in spin-integrated ARPES.

In order to get more insight into the theoretical aspects of the relation between the magnetic configuration and the electronic structure, we have employed the {\it ab initio} calculations. EuO band structure calculations for FM, and two antiferromagnetic phases,  AFM-I, and AFM-II \cite{Schlipf2013,Miyazaki2010}, visualize the band renormalizations under various magnetic arrangements of Eu $4f$ moments.  In addition, we analyze the role of spin-orbit coupling (SOC), which leads to a directional dependence of band energies. We find this effect to be particularly important for the Eu $4f$ levels.


Our spin-ARPES spectra directly demonstrate the exchange splitting in O $2p$, showing two exchange split doublets, as expected from the band structure calculations. The splitting vanishes when approaching $T_C$, as expected from the theoretical models \cite{Fedorov2021,Schiller2001c}. Furthermore, we show a difference in the Eu $4f$ spectral shape in spin-up and spin-down channels which we ascribe to the hybridization with the spin-up O $2p$ levels.

EuO is a highly reactive material, posing significant experimental challenges even when keeping the highest standards of the ultra-high vacuum methodology.  Spin-integrated data throughout $T_C$ demonstrates that sample aging effects in particular affect the O $2p$ bands, preventing unambiguous analysis of influence of magnetic phase transition on their spectral shape.





\begin{figure}
		\includegraphics[width=1\columnwidth]{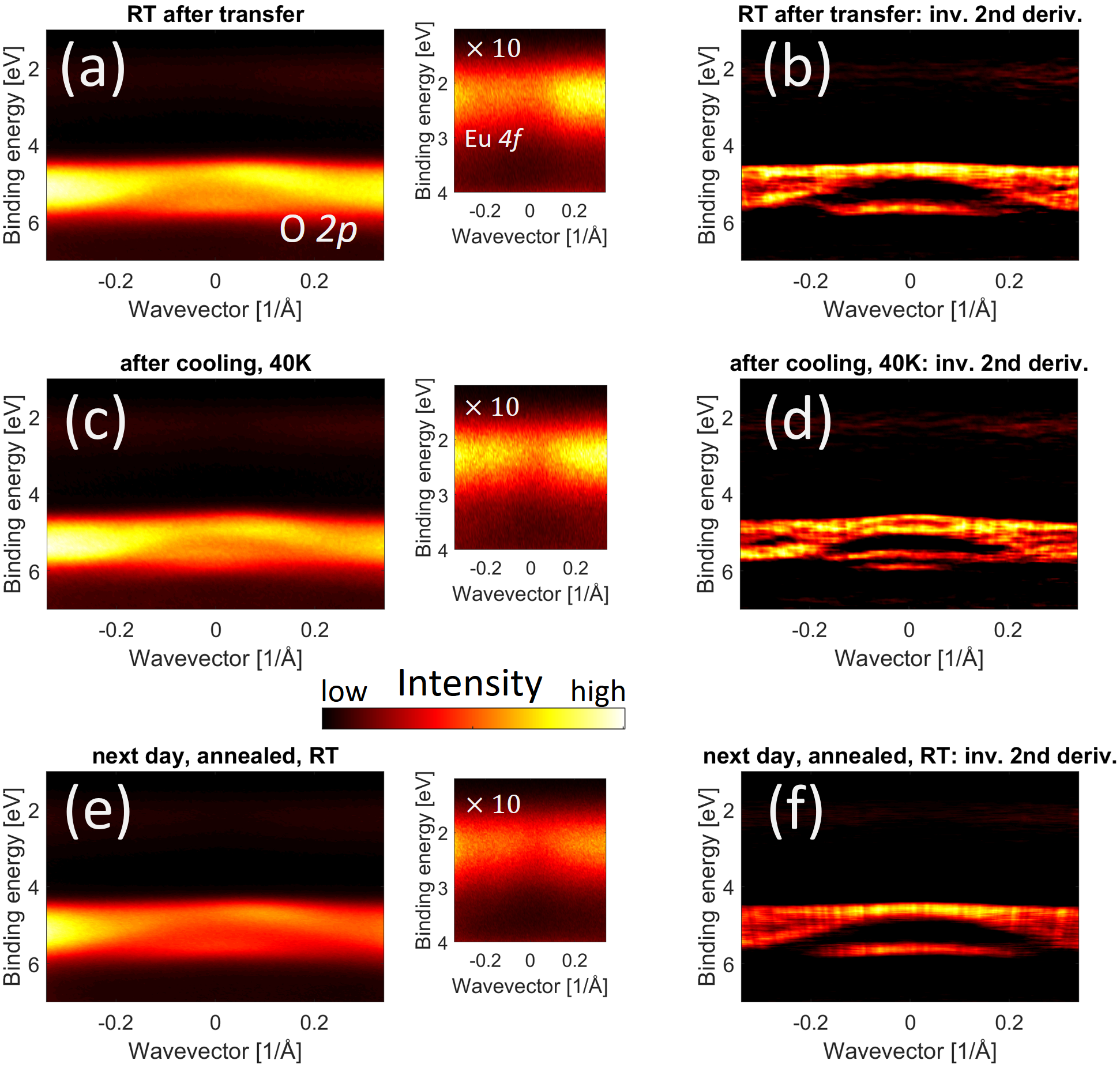}
	\caption{High-resolution ARPES spectra from "Sample A" taken at RT immediately after the vacuum transfer and sample preparation, immediately after the cooling to 40 K, and next day after another preparation cycle. Left column shows E(k) maps and right column their inverted second derivatives. Middle columns replot the E(k) maps in the Eu $4f$ regions normalized $\times 10$. (a)-(b) RT spectra after sample transfer and a preparation cycle. (c)-(d) Spectra after cooling the sample to $\sim 40$ K. (e)-(f) RT spectra taken next day after another preparation cycle. }
	\label{LT_and_RT}
\end{figure}

\begin{figure*}
		\centering
		\includegraphics[width=1.8\columnwidth]{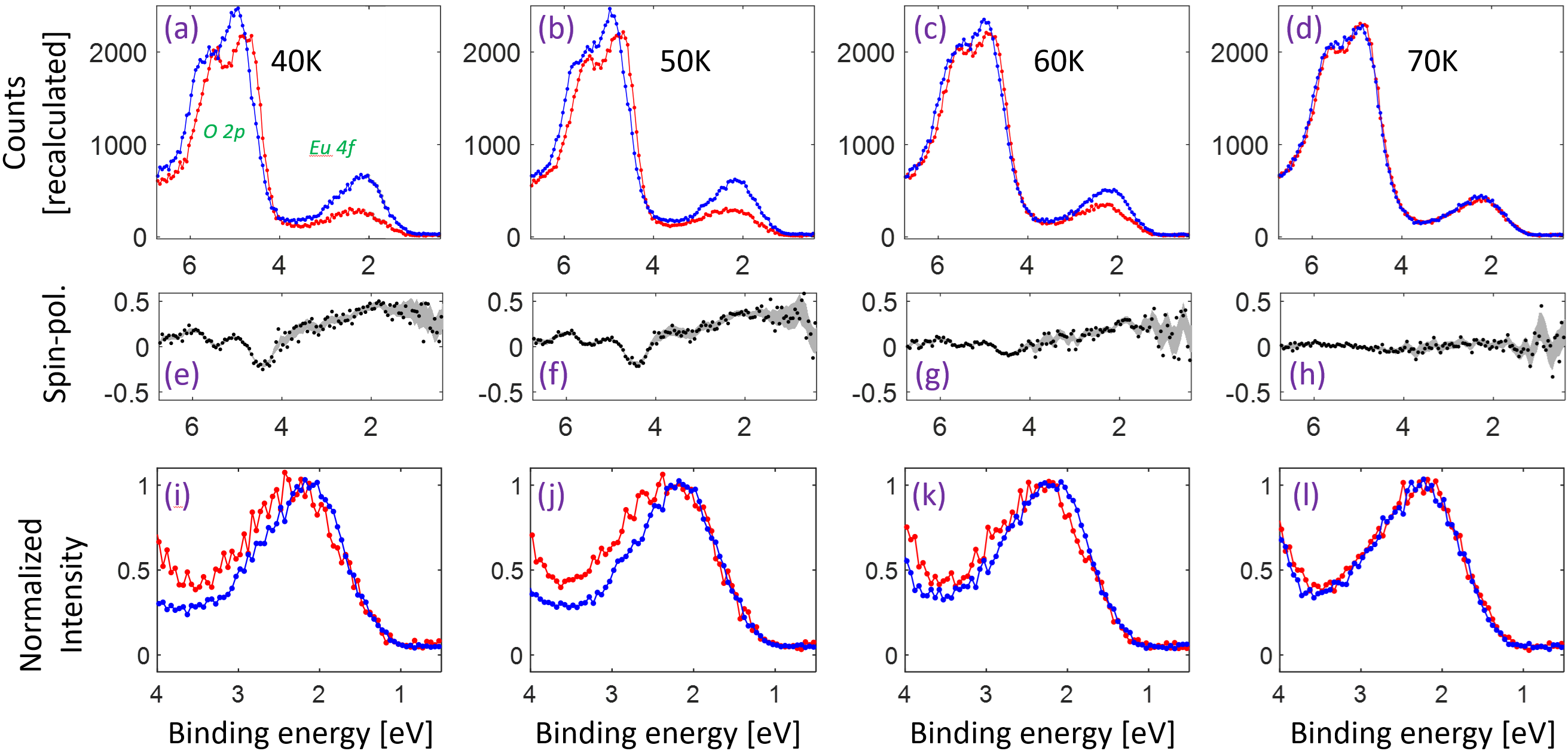}
	\caption{Temperature-dependent spin-polarized photoemission of Eu 4$f$ (located at $\sim2\,$eV) and O 2$p$ (at $\sim5\,$eV) bands taken on "Sample A". (a-d): data from the spin-up (blue) and spin-down spectrum (red) of the spin-detector corrected by the Sherman function S = 0.29. (e-f): respective calculated spin-polarizations, gray areas depict standard deviations for spin polarization curves. (i-l): spectra from (a-d) magnified in the Eu $4f$ region and normalized separately for up/down spectra. Temperatures from left to right: {40}{K}, {50}{K}, {60}{K}, {70}{K}, temperature calibration accuracy $\sim \pm 10$ K.}
	\label{ferrum}
\end{figure*}

\begin{figure}
		\centering
		\includegraphics[width=1\columnwidth]{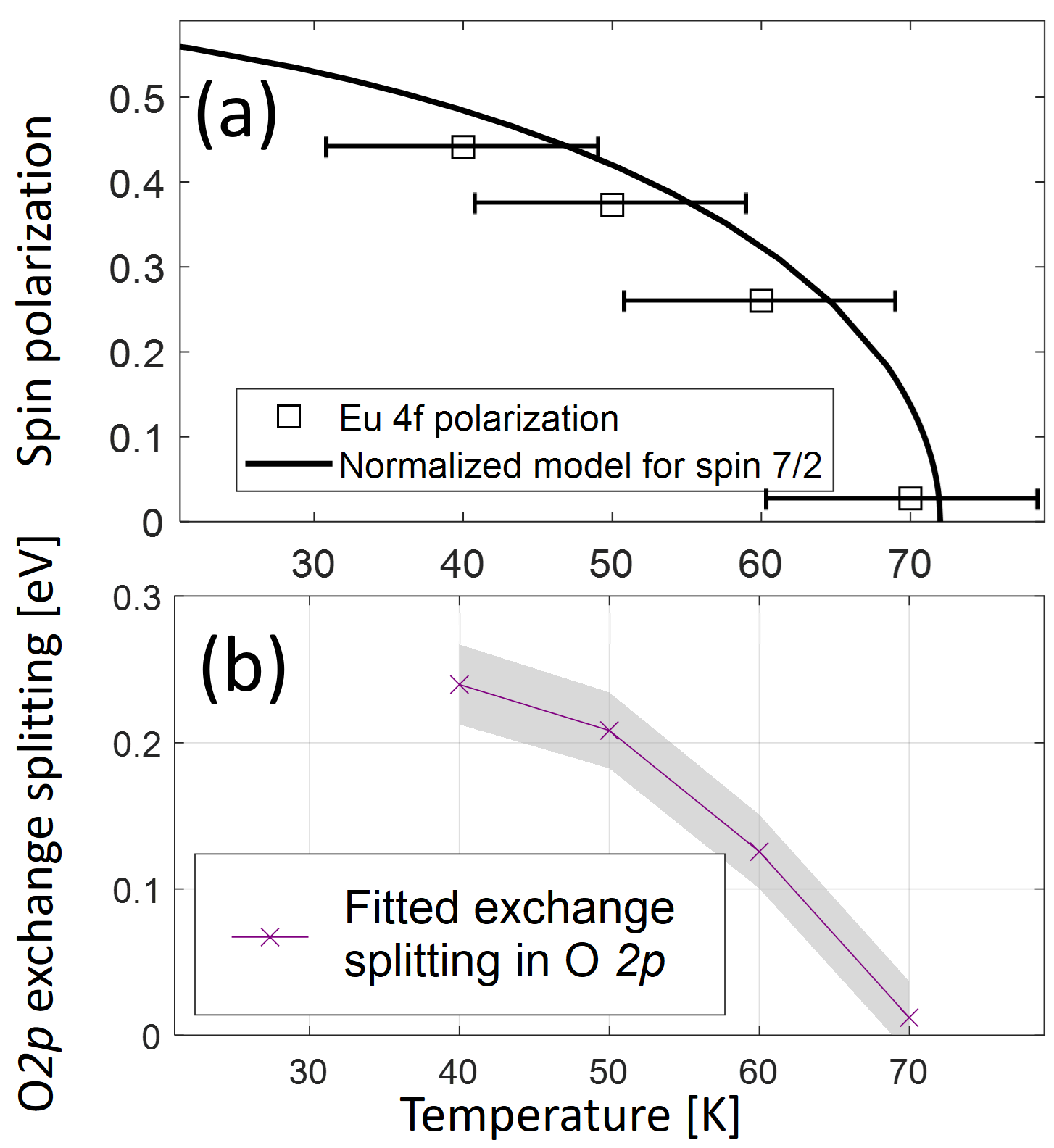}
	\caption{(a) Temperature dependence of the spin-polarization as obtained by integrating the Eu 4$f$ peak of the spectra from "Sample A" shown in Figs.\,\ref{ferrum}(a-d). The temperature calibration accuracy of $\pm 10$ K is indicated with error bars. (b) O 2$p$ exchange-splitting as obtained by fitting the data of Fig. \ref{ferrum}(a-d) (crosses) and of Fig. \ref{ferrum}(i-k) (circles). See the text and Supplemental Material Fig. S7 for details.}
	\label{fit_results}
\end{figure}

\begin{figure}
		\includegraphics[width=0.8\columnwidth]{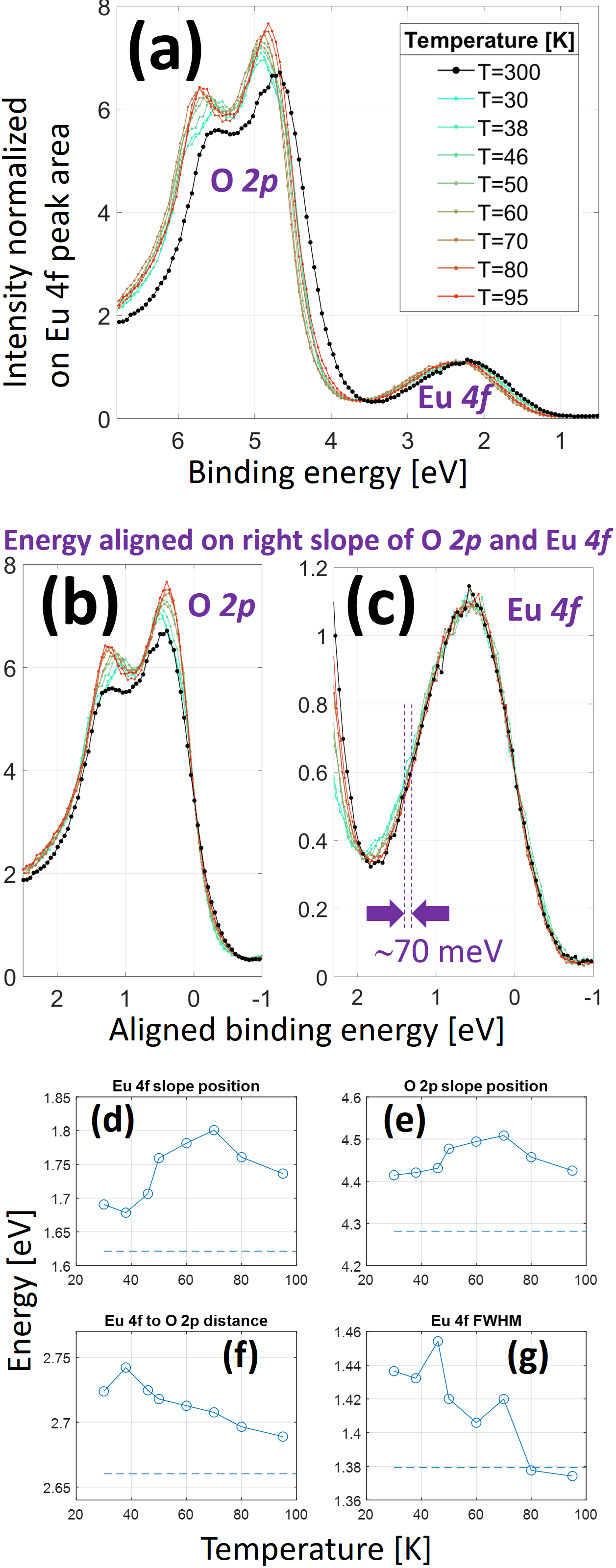}
	\caption{(a): Set of temperature-dependent normal emission spectra normalized on the Eu $4f$ area from "Sample B". The legend shows chronological sequence of taking the spectra. (b): O $2p$ peak area from (a) with the same intensity scaling, but with binding energies aligned on the right (lower binding energy) slope at half-maximum of 300K spectrum to allow comparison of the peaks widths and shapes. (c): the same as (b) but for Eu $4f$. Temperature dependent positions of the Eu $4f$ and O $2p$ slopes are plotted in (d) and (e), while (f) shows their difference, the distance between the two peaks. (g): FWHM of the Eu $4f$, taken on slightly smoothed spectra. Dashed lines in (d-g) show these values for 300K.}
	\label{LT_and_RT_Tristan}
\end{figure}

\section{Experimental results}

EuO thin films were grown in a molecular beam epitaxy (MBE) system with a residual gas pressure $p < 2 \times 10^{-10}$ mbar on a Cu(001) single crystal substrate. The films are epitaxial with EuO(001)$||$Cu(001), and $d=25$\,nm thickness resulting in a bulk-like electronic structure \cite{Prinz2016}. The LEED and RHEED patterns obtained on the EuO(001) films can be found in the Supplemental Material, Fig. S1. Stoichiometry of the EuO films was achieved by using the distillation method \cite{Ulbricht2008, Sutarto2009, Caspers2013, Loemker2019,Rosenberger2022} and was confirmed by in-situ X-ray photoelectron spectroscopy (XPS) (Supplemental Material, Fig. S2). The spin polarization of the photoemitted electrons was reversed after the sample was remagnetized, confirming its ferromagnetic character (Supplemental Material, Fig. S3). The bulk-like Curie temperature of $T_C \simeq 72$ K, was verified using vibrating sample magnetometry (VSM) measurements (Supplemental Material, Fig. S4). High-resolution ARPES and spin-ARPES measurements were carried out in another ultra-high-vacuum (UHV) chamber ($p < 1 \times 10^{-10}$ mbar). Samples were transferred using a transportable UHV shuttle. After the vacuum transfer, the samples were annealed up to $300^\circ$C for 2 min to desorb surface contaminants. The spectra were taken using non-monochromatized He I$\alpha$ resonance radiation with $h\nu=21.22$ eV. The detailed discussion of the possible electronic transitions within the free electron final state model is presented in the Supplemental Information, Fig. S5. The energy resolution in the ARPES mode was $20$ meV and in the spin-ARPES mode 50 meV. The parallel momentum resolution in spin-ARPES mode was $\leq 0.1$\AA$^{-1}$.  The spin-polarization of the photoelectrons was measured at normal emission with a FERRUM spin detector \cite{Escher2011} that has a Sherman function $S = 0.29$. The FERRUM detector measures two spectra $I_+$ and $I_-$ for the target magnetized in opposite directions. Spin polarization is calculated as $P = (1/S)\frac{I_+ - I_-}{I_+ + I _-}$, and the up/down spectra are reconstructed as $I_{\uparrow(\downarrow)} = 0.5 (I_+ + I _-)(1\pm P)$.

We have measured two EuO films grown on different days, which we denote the "Sample A" and the "Sample B". Measuring EuO surfaces with laterally averaging technique such as ARPES is challenging due to the EuO surface reactivity and surface aging. Because of this, even though the ultra-high vacuum transfer between MBE and ARPES chambers has been made immediately after the growth, flashing of the sample to $300^\circ$C before the measurement was needed to observe sharply dispersing bands.


{\it Angle-resolved and spin-integrated photoemission.---}Figure \ref{LT_and_RT} compares room temperature (RT) and 40 K spin-integrated ARPES maps from "Sample A" ("Sample B" shows nearly identical results). Chronologically, after the vacuum transfer and preparation by annealing, the RT spectrum in Fig. \ref{LT_and_RT}(a) was measured. Immediately after cooling to 40 K the spectrum in (c) was measured, and a further RT spectrum (e) was measured the next day after warming up the cryostat and another annealing cycle. Second derivative spectra indicate that near normal emission the O $2p$ exhibits two-peak structure at RT and four-peak structure at 40 K, in agreement with similar analysis used in the previous work  \cite{Miyazaki2009}.


{\it Angle-resolved and spin-resolved photoemission.---}Figure \ref{ferrum} shows a series of temperature-dependent spin-ARPES spectra from "Sample A" (a similar set of spectra for "Sample B" is presented in Supplemental Material Fig. S8). Figure \ref{ferrum}(a) shows spin-resolved spectra for the sample in a ferromagnetic state at $T \simeq 40$ K, where the temperature-derived depolarization should be small. The O $2p$ manifold in Fig. \ref{ferrum}(a) shows the exchange splitting of $0.23$ eV for a more pronounced lower binding energy peak in the doublet. Figures \ref{ferrum}(a)-(d) reveal how the spin-resolved spectra evolve with the increasing temperature and (e)-(h) present the corresponding spin polarization distributions. It is evident that the spin-polarization in the photoemitted ensemble converges to 0 upon approaching $T_C \simeq 72$ K (Fig. \ref{ferrum}(d) and (h)). Details of fitting the spin-polarized O $2p$ features with Voigt doublets are presented in Supplemental Material Fig. S7.

Due to the intra-atomic exchange splitting of the order of $\sim$10 eV the Eu $4f$ manifold is virtually 100 \% spin polarized (see Supplemental Fig. S9), however, in Fig. \ref{ferrum}(e) the Eu 4$f$ band shows a spin-polarization of only $\sim 50\%$. The reason is that the sample has been magnetized along one of the <111> easy axes, therefore only the projection onto the in-plane [010] direction (the detector quantization axis) is probed, and the measured polarization is reduced by $1/\sqrt{3}$, i.e. down to $\sim 58\%$. The large exchange splitting of $\sim$10 eV of Eu $4f$ bands results from the $4f$ local moments and does not require any magnetic order. Figure \ref{fit_results}(a) summarizes the temperature-dependent spin polarization in Eu $4f$ levels of Fig. \ref{ferrum}(a-d).

The magnitude of the Eu $4f$ local moment is large and therefore virtually temperature independent, however, the spectral shape of Eu $4f$ features in up/down spin-ARPES spectra may differ if they are affected by the hybridization with the spin-up portion of O $2p$. In Fig. \ref{ferrum}(i-l) we plot the Eu $4f$ portion of up/down spectra from (a-d), but separately normalized to visualize the differences in their spectral shapes which are discussed in the later part of the manuscript by comparison to the theoretical calculations.

Figure \ref{fit_results}(b) presents the values of the O $2p$ exchange splitting extracted from the fitted spectra. The crosses mark points obtained following the analysis of the spin-resolved spectra using the Sherman function of S=0.29 showing that the exchange splitting approaches zero when the temperature reaches $T_C$. The splitting in Fig. \ref{ferrum}(a) of 0.23 eV is in agreement with previous spin-integrated ARPES work \cite{Miyazaki2009} where 0.21 eV was reported. We note, however, that the actual exchange splitting in the O $2p$ is likely slightly larger, since our analysis does not include the correction for the <111> easy axis projection on the detector [010] quantization axis.


{\it Temperature-dependent peak widths.---}Figure \ref{LT_and_RT_Tristan} shows the temperature-dependent set of normal emission spectra from "Sample B". Fig. \ref{LT_and_RT_Tristan}(a) shows the temperature series of normal emission spectra normalized on the Eu $4f$ peak. Chronologically, the 300K data was measured first, then the sample was cooled down to $\sim 30$ K, and warmed up in steps as indicated in the legend of Fig. \ref{LT_and_RT_Tristan}(a). The intensity ratio between O $2p$ and Eu $4f$ steadily grows over time, that is, depends mostly on the surface aging and not on the temperature. Therefore, the changes in the shape of the O $2p$ peak in Fig. \ref{LT_and_RT_Tristan}(b) cannot be considered as being of a purely magnetic origin. Eu $4f$ spectra in Fig. \ref{LT_and_RT_Tristan}(c) exhibit changes in the FWHM of the order of 70 meV, summarized in (g), with the general trend of the FWHM getting narrower at higher temperatures which suggests the bandwidth narrowing with the increased magnetic disorder, in agreement with the renormalization in spin polarized spectra shown in Fig. \ref{ferrum}(i-l). This picture also agrees with the results on EuS \cite{Fedorov2021}.



Figs. \ref{LT_and_RT_Tristan}(a) and (d-e) indicate binding energy shifts between the spectra, with the similar order of magnitude but different behavior as compared to the previous work \cite{Miyazaki2009}. The origin of this difference might be related either to the surface aging or to the band bending and the photovoltage between the Cu(001) and the EuO film. Such effects are difficult to quantify, however, the separation of the peaks in Fig. \ref{LT_and_RT_Tristan}(f) is consistent with the Ref. \cite{Miyazaki2009}, being larger for the FM phase.



\begin{figure*}
		\centering
		\includegraphics[width=1.8\columnwidth]{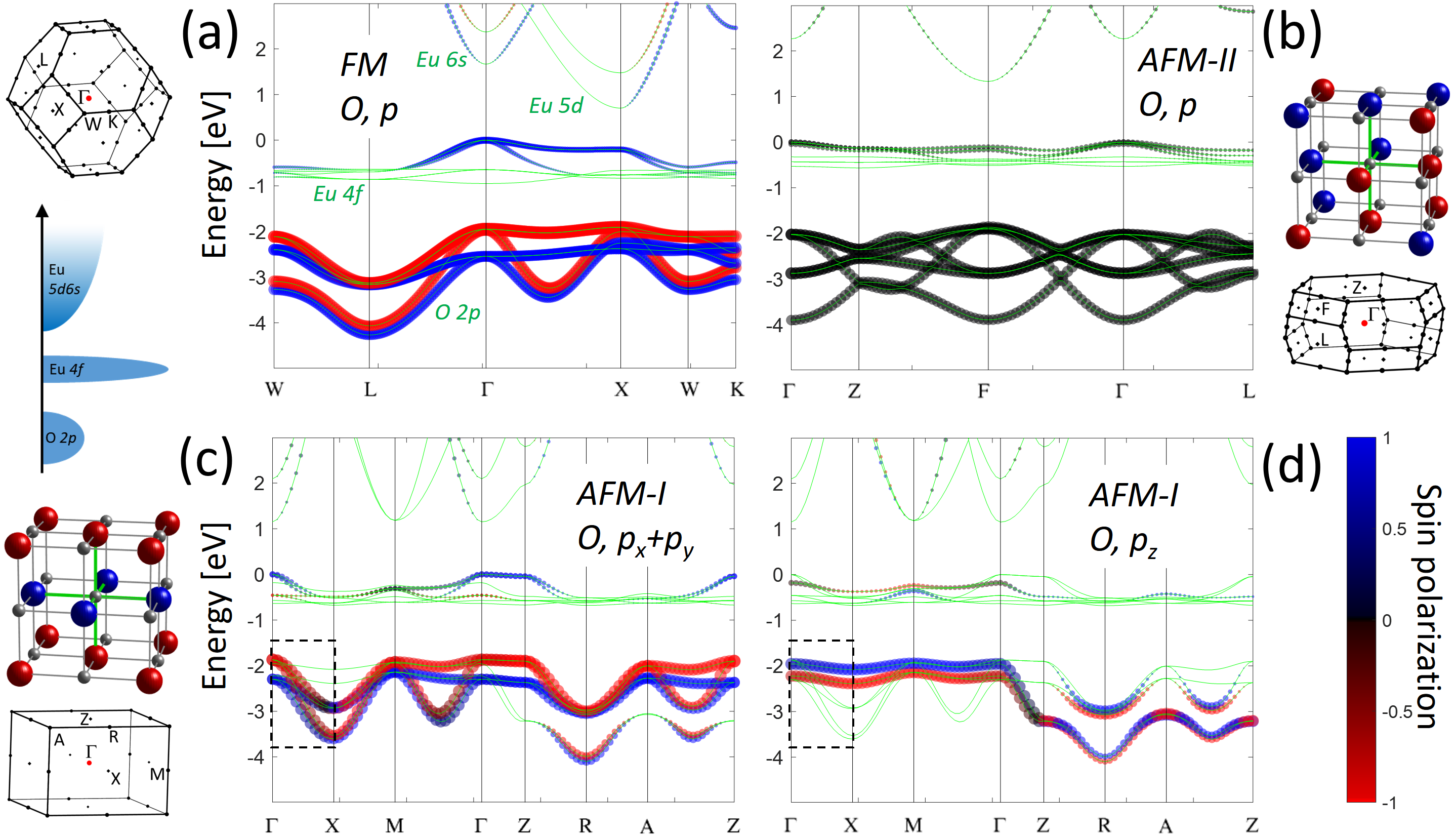}
	\caption{DFT+$U$ band structure and the corresponding BZs of (a) FM, (b) AFM-II and (c)-(d) AFM-I phases of EuO. Energy eigenvalues are depicted with green lines. The size of colored filled circles indicate the $p$ orbital character of the oxygen atom, and colors depict the spin-polarization in these orbital projections. Lattice spin up/dn configurations of AFM-II and AFM-I phases are shown next to panels (b) and (c) respectively. The two regions marked by dashed line boxes in (c) and (d) indicate a reversal of the spin character for differently dispersing bands.}
	\label{FMAFM}
\end{figure*}

\section{Theoretical calculations}



{\it EuO band structure in FM, AFM-I, and AFM-II phases.---}Realistic DFT+$U$ bulk band structures of EuO in FM and AFM phases are presented in Fig. \ref{FMAFM}. These calculations were performed using the full-potential linearized augmented plane wave method as implemented in the WIEN2k code \cite{Blaha2001} at the cubic lattice constant of 5.14 \AA. For the exchange-correlation functional we used the generalized gradient approximation (GGA) \cite{Perdew1996}. In order to correctly describe the relative positions of bands as measured in photoemission experiments, we have included static local electronic correlations  to the GGA potential in the GGA+$U$ method \cite{Larson2006} with $U = 7.9$\,eV on Eu $4f$ orbitals. The use of bulk band structure for interpreting the ARPES data is justified since no well-defined surface states are expected in the Eu $4f$ and O $2p$ manifolds for the EuO(001) surface \cite{Klinkhammer2014}.

Note that in Fig. \ref{FMAFM} we have used an initial energy scale, that is negative for occupied bands and referenced to the top of the $4f$ manifold, and in Figs. \ref{LT_and_RT}, \ref{ferrum}, and \ref{LT_and_RT_Tristan} we use a binding energy scale referenced to the Fermi level measured on the metal plate in contact with the sample. The two scales are shifted with respect to each other, and similarly to Ref. \cite{Riley2018} the theoretical energy separation between $4f$ and $2p$ manifold is too small, see also Fig. S5(a)-(b) of the Supplemental Material. This discrepancy does not influence our conclusions, and has been previously addressed by calculations with correlation effects included from first principles \cite{Schlipf2013}.

The symbol size in Fig. \ref{FMAFM} indicates the local O $2p$ character on a particular site, while blue (red) color mark the predominantly majority (minority) spin character. The calculated cases are: the FM phase with all Eu spins aligned, the AFM-I with 4 Eu nearest neighbours (NN) and 2 NN in opposite spin directions, and the AFM-II where 3 NN Eu are spin-up and the other 3 spin-down \cite{Miyazaki2010}.
The insets in Fig. \ref{FMAFM} show schematics of these magnetic configurations with bigger blue (red) balls as Eu majority (minority) ions, and small balls as oxygen ions. In addition, the Supplemental Material Fig. S10 visualizes the influence of the FM and AFM-I phases on the spectral shapes and bandwidths by directly comparing the bands along the same reciprocal directions, as referred to the crystal lattice, for the two phases. Detailed plots of various orbital characters and additional comparisons are provided in the Supplemental Material Fig. S9, and the picture that emerges reveals a complex character of energy splittings in the O $2p$ manifold; the exchange splitting depends not only on the BZ direction but also on the orbital character of the electronic states. In particular, in the AFM-I phase, locally, the O $2p$ orbitals are partially polarized, and when looking at a particular O site in the lattice, the polarization order reverses depending on the orbital character. Without the loss of generality, let us assume that the intralayer FM arrangement is within $x$-$y$ planes, see a schematic crystal lattice next to Fig.  \ref{FMAFM}(c). Along $z$, subsequent layers are arranged antiparallel. Therefore, despite the fact that O $2p$ are locally polarized their global polarization of course vanishes. Then, O $2p_x$ and $2p_y$ interact primarily with Eu $4f$ moments within the same plane, and O $2p_z$ with Eu $4f$ moments from adjacent planes. In the FM case the $p_x, p_y$ and $p_z$ orbital character is mixed for all bands with the O $2p$ dispersion reminiscent of the valence bands of the classical semiconductors \cite{Yu2010}, but for AFM-I along many high-symmetry directions O $2p$ bands split, having either $p_x + p_y$ or $p_z$ orbital character (suggesting $\pi$- and $\sigma$-like bonding for AFM phases). This leads to the spin reversal in bands derived from in-plane and out-of-plane (in our geometry) O $2p$ orbitals. As an example we have highlighted bands along $\Gamma$-X by dashed line boxes in Fig. \ref{FMAFM}(c) and (d). The order of red-blue colors reverses between these two cases, in particular near the $\Gamma$ point, which means their spin polarizations are anti-parallel, with the local polarization reaching  98\% for the $p_{xy}$ orbital and 87\% for the $p_z$ orbital.

The understanding of the magnetism in Eu chalcogenides has been based on the various indirect exchange mechanisms \cite{Steeneken2002,Kasuya1972,Mauger1986}. Establishing the values of NN and next NN exchange integrals $J1$ and $J2$, is still an active research area \cite{Wan2011,Sollinger2010,Liu2012}. For instance, Wan {\it et al.} \cite{Wan2011} have discussed a considerable hybridization between Eu $4f$ and O $2p$ leading to the superexchange, which, however, turns out to be a second order effect compared to other exchange mechanisms. Consistently, we notice a hybridization between O $2p$ and Eu $4f$ manifolds. In particular near the $\Gamma$ points a partial $p$ character of the bands within the Eu $4f$ manifold is observed, see Fig. \ref{FMAFM}(a) and Supplementary Fig. S9. Accordingly, $f$ character is observed in spin-up part of the O $2p$ manifold (Supplementary Fig. S9), revealing the nature of O $2p$ - Eu $4f$ hybridization. This hybridization leads to the increased bandwidth of the Eu $4f$ manifold in the FM phase, in agreement with the results on EuS \cite{Fedorov2021}, and is responsible for the renormalization of the spectral weight in spin-ARPES from the Eu $4f$, Fig. \ref{ferrum}(i-l). Our theoretical bandwidths in eV for FM, AFM-I, and AFM-II for Eu $4f$ are 0.95, 0.68, and 0.57, and for O $2p$ are 2.32, 2.23, 2.03 respectively, indicating the Pauli localization in the AFM phases. Notably, the hybridization of Eu $4f$ with the conduction bands is less pronounced (Supplementary Fig. S9), suggesting it has less influence on the Eu $4f$ bandwidth.

Also the Eu $5d$ orbitals slightly hybridize with Eu $4f$, with the difference in the $5d$ orbital shapes that despite being centered on Eu sites can overlap with both their own on-site $4f$ and some of their 12 NN $4f$, depending on the $d$ orbital symmetry. The non-vanishing local spin polarization in the Eu $5d$ also for the AFM-II phase (Supplemental Material Fig. S9) seems to be related to this overlap.

\begin{figure}
		\centering
		\includegraphics[width=5cm]{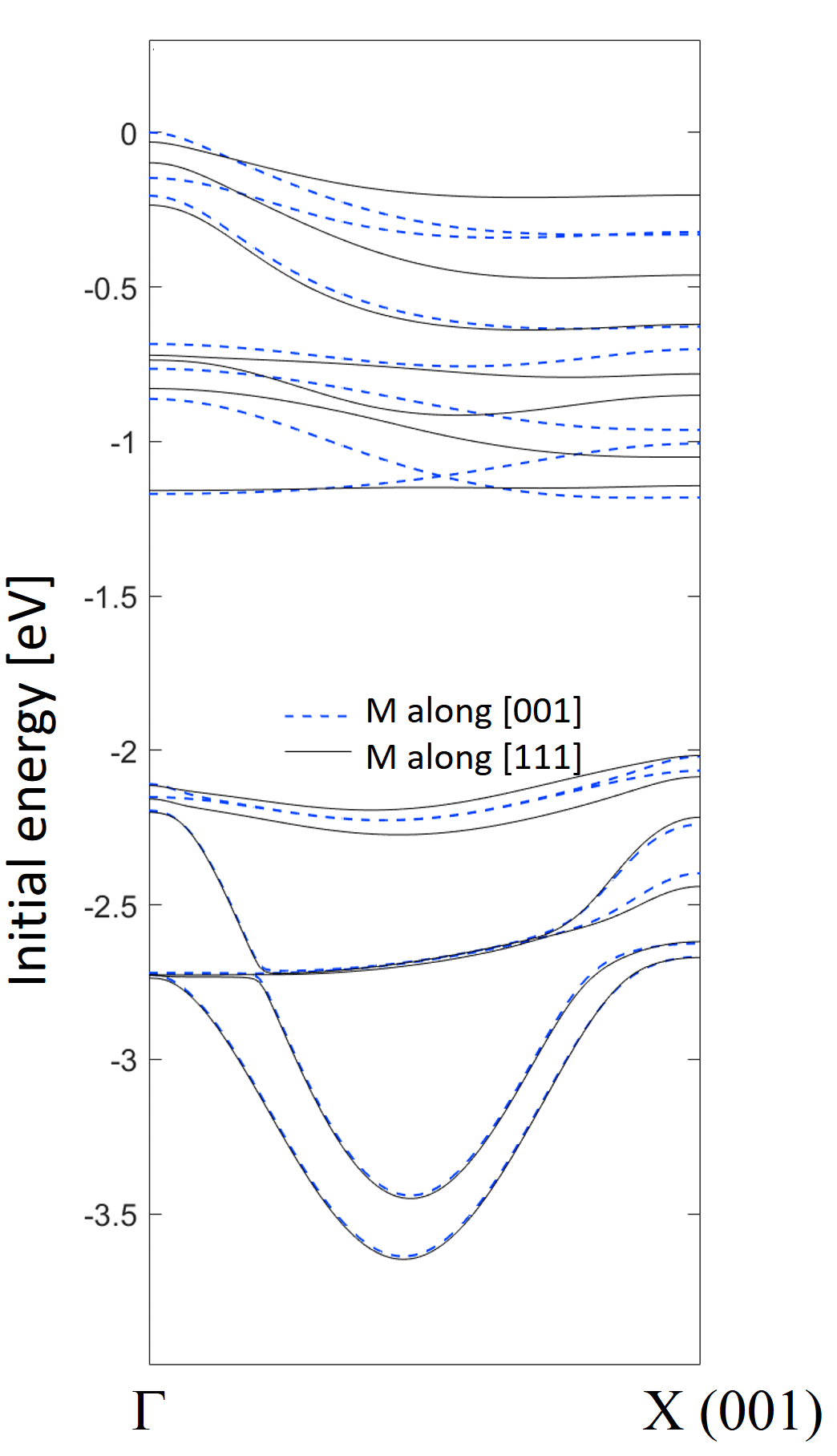}
	\caption{Theoretical DFT calculations with SOC included. (a) The band structure along the [001] $\Gamma-X$ direction (in our case the direction normal to the surface), for the magnetization $M$ along [111] (solid line) and [001] (dashed line).}
	\label{DFT_SOC}
\end{figure}


{\it Band renormalizations due to spin-orbit coupling.---}Typically the effects of spin-orbit coupling (SOC) are not considered in discussing the band structure of EuO. While the easy axis in our films is along [111], it is interesting to establish how would the band dispersions renormalize in case of a sample hypothetically magnetized along another direction. Our calculations were performed using WIEN2k with SOC included using the second-variational approach \cite{Blaha2001}. Our spin-ARPES spectra were taken at normal emission, which for our geometry probes the [001] $\Gamma-X$ direction in the BZ due to the conservation of the parallel momentum. Therefore, we focus on the band structure along this direction, with the result for two different magnetization directions presented in Fig. \ref{DFT_SOC}. The solid lines and dashed lines in Fig. \ref{DFT_SOC} represent the band structure along $\Gamma-X$, with SOC included, for the sample magnetized along [111] and [001] respectively. One can see that changing the magnetization direction renormalizes band dispersions, with lifted degeneracies and shifts of the order of 50-200 meV in the Eu $4f$ manifold, while the effects in O $2p$ are much smaller.

\section{Discussion}

The vanishing temperature-dependent exchange splitting in the $p$ manifold has been previously conjectured from the spin-integrated ARPES data for EuO \cite{Miyazaki2009} and EuS \cite{Fedorov2021}. Our data in Fig. \ref{ferrum} directly demonstrates this effect in O $2p$ in spin-ARPES from remanently magnetized EuO thin films. It also clarifies earlier spin-polarized investigation of EuO, where the reverse ordering in the exchange splitting in O $2p$ has been proposed from analysis of the spin-polarized spectra at $h\nu=135$ eV. The Eu $4f$ bands, which originate form the $4f$ LMs and for which the exchange splitting is virtually temperature-independent, are strongly polarized in the FM state, and their polarization vanishes at the $T_C$. Therefore they can serve as a gauge of the global magnetization of the system. On the other hand, as shown in Fig. \ref{ferrum}(i-l), the spectral shape of Eu $4f$ features is not the same in up/down channels. The Eu $4f$ spin-up (majority) spectrum, depicted in blue color in Figs. \ref{ferrum}(i-k), is slightly shifted towards lower binding energies, and the spin-down (minority, red color) exhibits a tail on the high binding energy side. Theoretical calculations in Fig. \ref{FMAFM}(a) and Supplemental Material Fig. S9 show that these effects are related to hybridization between Eu $4f$ and O $2p$, which is now demonstrated experimentally. In particular, the dispersive lower binding energy portion of the Eu $4f$ has an admixture of O $2p$, and the flat portion of the spin-up part of the O $2p$ has an admixture of Eu $4f$. Similar conclusions were made for EuS \cite{Fedorov2021}.

Our AFM calculations in Fig. \ref{FMAFM} (c-d) show how the induced moment on O $2p$ does not vanish in the AFM-I phase and how it depends on the configuration of the 6 NN Eu $4f$ moments. This suggests that also in the fully disordered scenario the magnitude of the induced local moment on O $2p$ will not vanish, if it is taken as a normalized sum of the 6 NN (fully disordered) Eu $4f$ moments. However, since the FM exchange splitting in O $2p$ is  $\sim 0.23$ eV, and the bandwidth of the O $2p$ manifold of the order of 2 eV, no signatures of this residual O $2p$ local moments are expected in the dispersive O $2p$ band manifold in the PM phase, as recently simulated theoretically in case of S $3p$ states in EuS. It is possible that in the vicinity of $T_C$ non-vanishing spin-correlation length exists in the PM phase \cite{Boeni1986}, however, in transition metals estimations of non-vanishing correlation lengths above $T_C$ are of the order of 2-3 atomic distances \cite{Melnikov2019}, while EuO exhibits significant phonon modes softening when approaching $T_C$ \cite{Passell1976}, therefore likely making all phonon modes active in the vicinity of $T_C$.

In addition to these effects, our calculations in Fig. \ref{DFT_SOC} show that sizeable directional effects of SOC are present in particular within the Eu $4f$ manifold. So far the temperature-dependent calculations \cite{Schiller2001c,Fedorov2021} did not include SOC, and one can expect non-negligible additional renormalizations of the Eu $4f$ spectral weight between FM and PM phases when SOC is included.










\section{Summary}



Through the temperature-dependent spin-ARPES spectra we have directly demonstrated the exchange splitting of the O $2p$ manifold in remanently magnetized EuO(001) thin films. The exchange splitting vanishes when approaching $T_C$, consistent for the expectations for the PM phase. Small renormalization of the spectral shape between spin up and down spectral region related to Eu $4f$ is also observed and ascribed to the hybridization with O $2p$.

Spin-integrated experimental spectra show dispersive features in the O $2p$ manifold indicating clean and well-ordered EuO(001) surfaces. Analysis of the temperature-dependent peak widths in the spin-integrated spectra is obscured by the sample aging effects, in particular for the O $2p$ manifold. The Eu $4f$ spectral region is less affected, and exhibits temperature-dependent narrowing at higher temperatures, ascribed to the spectral narrowing in the magnetically disordered phases, in agreement with the predictions for EuS \cite{Fedorov2021}.

Our DFT calculations with a Hubbard $U$ term for FM, AFM-I, and AFM-II phases reveal a complex nature of the exchange splitting in the O $2p$ manifold, which depends on the arrangement of the 6 NN Eu $4f$ local moments. The bandwidths of the Eu $4f$ and O $2p$ manifolds decrease in AFM phases, indicating the expected Pauli localization. Calculations including SOC reveal renormalizations of the band dispersions which are particularly strong in Eu $4f$ bands and may additionally influence the interpretation of their temperature dependent spectral shape.

\section{Acknowledgements}

Authors would like to thank G. Bihlmayer, M. Donath, Y. Mokrousov, and M. dos Santos for fruitful discussions. We thank O. Petracic and the J{\"u}lich Centre for Neutron Science for providing measurement time at the magnetometers. M.M. acknowledges funding by the Deutsche Forschungsgemeinschaft (DFG, German Research Foundation) "SFB 1432" Project-ID 425217212. L.P. and T.H. acknowledge funding by the German Research Foundation (DFG) priority program SPP1666 {\it Topological Insulators: Materials - Fundamental Properties - Devices} under the number PL 712/2-1. R.P. and A.K. acknowledge funding by the DFG within CRC/TRR80 (project number 107745057, subproject G3) and CRC1242 (project number 278162697, subproject C02) and computational time at the Leibniz Rechenzentrum Garching, project pr87ro.


%

\end{document}